\documentstyle[prb,aps]{revtex}
\begin{document}
\draft
\title{Behavior of excitonic levels in symmetric and asymmetric
double quantum wells in a magnetic field}
\author{Francisco Vera and Zdenka Barticevic}
\address{Universidad T\'{e}cnica Federico Santa Mar\'{\i}a,
casilla 110-V Valpara\'{\i}so, Chile}
\date{\today}
\maketitle
\begin{abstract}
We studied theoretically the
excitonic energy levels and the optical absorption spectra
for double quantum wells,both symmetric and asymmetric,
in the presence of an homogeneous magnetic field.
Within the effective mass approach, we expanded the
excitonic wave-function, in an orthogonal basis formed by
products of electron and hole wave-functions
in the growth direction $z$, and one particle
solutions of the magnetic Hamiltonian in the $x-y$ plane.
We applied our method to the case of $Al_xGa_{1-x}As$,
for which we showed how the exciton wave-function vary,
and how the basis functions are mixed in a non trivial way
by the effect of the Coulomb potential.
By taking into account all the mixing between the elements
in our base, we get anti-crossings between
excited excitonic states not reported previously.
\end{abstract}


\section{Introduction\protect\\}
\label{sec:Introduction}

In double quantum wells, excitons
are more complex than in a single quantum well, because the
electron and hole wave-functions can be localized either
in the same well, or in different wells, or in both.
The asymmetry of the quantum well produces
transitions which are prohibited in the optical spectrum,
of the symmetric case.

The theoretical study of excitons in double quantum wells,
began with the work of Kamizato and Matsuura \cite{kamizato}.
They studied the excitonic levels and binding energy
of symmetric wells
as a function of the wells and barrier widths.
Dignam and Sipe \cite{dignam} improved the calculation for the
excitonic levels behavior for wide barriers, incorporating
the Coulomb interaction between excitonic states.
Their work included both symmetric and asymmetric wells,
and the presence of an external electric field
(see also Ref.\ \onlinecite{ferreira1990}),
but their method of calculation is not suitable for thin barriers.
Cen and Bajaj \cite{cen} improved the variational
method used in Ref.\ \onlinecite{kamizato}, obtaining the
exciton binding energy, in symmetric wells in an external magnetic
field, for arbitrary wells and barrier width.
Recently Dzyubenko and Yablonskii \cite{dzyubenko},
using a non variational method,
studied the excitonic levels
for symmetric and almost symmetric double quantum wells,
as a function of the electric and magnetic field strength.
For an experimental study of excitonic levels
in $Al_xGa_{1-x}As$ symmetric double quantum wells
see Bayer $et\ al$ \cite{bayer}, and
for a recent review in the subject of excitons in
double quantum wells, see Ref.\ \onlinecite{ferreira1997}.

In this work, we studied the excitonic levels,
binding energy and optical transition probabilities,
in symmetric and asymmetric double quantum wells,
in the presence of a magnetic field,
as a function of barrier and well widths and
the magnetic field strength.
The interaction between light and heavy holes \cite{hernandez}
is not considered in our work
and we restrict the analysis only to excitons
composed by electrons and heavy holes.
Within the effective mass approach, we expand the
excitonic wave-function, in a orthogonal basis formed by
products of electron and hole wave-functions
in the growth direction $z$, and one particle
solutions of the magnetic Hamiltonian in the $x-y$ plane.

The Coulomb potential between electrons and holes
produces off-diagonal terms by mixing our basis states.
We obtained the energies and wave-functions
by diagonalizing the excitonic Hamiltonian
in a truncated basis.

In contrast to the majority of works in the literature
\cite{kamizato,dignam,cen,linnerud},
our basis set is orthogonal \cite{dzyubenko},
and we do not use a variational method.

We applied our method to study the first excitonic states,
in $GaAs$ ($Al_xGa_{1-x}As$) heterostructures.
For comparison purposes in the symmetric case,
we used the parameters of Ref.\ \onlinecite{cen}.
Our analysis of the basis states
involved in the excitonic wave-function
for different quantum wells geometries,
should be helpful in similar studies of
$III-V$ and $II-VI$ \cite{hernandez,ten,magnea} systems.

\section{Formalism\protect\\}
\label{sec:Formalism}

The effective mass Hamiltonian for excitons
in a double quantum well in the presence of
a magnetic field pointing towards $z$,
in the diagonal approximation \cite{bastard}
can be written as
\begin{equation}
\label{h-total}
H=H_0(z_e)+H_0(z_h)+H_{mag}(\rho)+V_{coul}(\rho,|z_e-z_h|).
\end{equation}
$H_0(z_e)$ is the 1-dimensional Hamiltonian for
electrons,
\begin{equation}
\label{h-ele}
H_0(z_e)=p_{ze}^2/2m_{ze}+V_e(z_e),
\end{equation}
$H_0(z_h)$ is the 1-dimensional Hamiltonian for
holes,
\begin{equation}
\label{h-hole}
H_0(z_h)=p_{zh}^2/2m_{zh}+V_h(z_h),
\end{equation}
and $V_e(z_e)$ ($V_h(z_h)$) is the potential that defines
the double quantum well for electrons (holes)
in the five regions of $z$, shown in Fig.\ \ref{pot-z}.
$H_{mag}(\rho)$ is the magnetic Hamiltonian in the
symmetric gauge, which depends on the relative coordinates of
electrons and holes in the $x-y$ plane,
\begin{equation}
\label{h-mag}
H_{mag}=\frac{(\vec{p}-q\vec{A})^2}{2\mu}+
\frac{qB}{m_{h,x-y}}l_z,
\end{equation}
where $\vec{p}$, $\mu$ and $m_{h,x-y}$ are the
momentum operator, reduced mass and hole mass, defined
in the $x-y$ plane.

$V_{coul}(\rho,|z_e-z_h|)$ is the Coulomb potential
between electrons and holes, including an effective
dielectric constant for the system.

We expanded the solution of the Hamiltonian \ref{h-total},
as a lineal combination of products of eigenfunctions of the
magnetic Hamiltonian in the $x-y$ plane (\ref{h-mag}),
and eigenfunctions of the electron and hole Hamiltonians
in the $z$ direction (\ref{h-ele} and \ref{h-hole}),
\begin{equation}
\label{expansion}
\Psi^{exc}_n=\sum_{\nu_\rho,\nu_e,\nu_h}C^{n}_{\nu_\rho,\nu_e,\nu_h}
\psi_{\nu_\rho}(\rho,\phi)\psi_{\nu_e}(z_e)\psi_{\nu_h}(z_h),
\end{equation}
in which
\begin{eqnarray}
\label{f-mag}
\psi_{\nu_{\rho},l}= &
 \frac{1}{2\pi}\left(\frac{2(n-l/2-|l|/2)g_B^{|l|+1}}
{(|l|/2+n-l/2)!}\right)^{1/2}e^{il\phi}(\frac{\rho}{i})^{|l|}\\
\nonumber
 &\times \ e^{-g_B\rho^2/2}L^{|l|}_{n-l/2-|l|/2}(g_B\rho^2),
\end{eqnarray}
where only $l=0$ functions are considered, and $g_B=\frac{qB}{2\hbar}$.
The electron wave-function
defined in the five regions of $z$, shown in Fig.\ \ref{pot-z},
is given by
\begin{equation}
\label{f-eh}
\psi(z_{e})= \left\{
\begin{array}{l}
a_1e^{k_1(z_e-z_1)} \\
a_2cos(k_2(z_e-z_1))+a_3sin(k_2(z_e-z_1)) \\
a_4e^{k_3(z_e-z_2)}+a_5e^{-k_3(z_e-z_2)} \\
a_6cos(k_4(z_e-z_3))+a_7sin(k_4(z_e-z_3)) \\
a_8e^{-k_5(z_e-z_4)}, \\
\end{array}
\right. 
\end{equation}
and the hole wave-function is given by similar expressions.

The Coulomb interaction
produces off-diagonal terms by mixing our basis states.
In order to obtain the system of equations for the
coefficients in the expansion (\ref{expansion}),
we need to evaluate the Coulomb integral
\begin{equation}
\label{i-total}
\int d\phi d\rho dz_e dz_h
\psi^{*}_{\nu'_\rho}\psi_{\nu'_e}\psi_{\nu'_h}
V_{coul}(\rho,|z_e-z_h|)\psi_{\nu_\rho}\psi_{\nu_e}\psi_{\nu_h}.
\end{equation}

The $\phi$ integral is trivial, because of $l_z$ conservation.
Using the explicit expansion of Laguerre Polynomials ($L_n$) in
$\psi_{\nu'_\rho}$ and $\psi_{\nu_\rho}$ (\ref{f-mag}),
the remaining of integral (\ref{i-total}) can be written
as a sum  of terms of the form:
\begin{equation}
\int_{-\infty}^{\infty} dz_edz_h
\psi_{\nu'_e}\psi_{\nu'_h}\psi_{\nu_e}\psi_{\nu_h}
\int_0^{\infty} \rho d\rho 
\frac{\rho^{2w}e^{-g_B\rho^2}}{\sqrt{\rho^2+(z_e-z_h)^2}}.
\end{equation}
By using
\begin{equation}
\frac{1}{\sqrt{\rho^2+(z_e-z_h)^2}}=\int_{0}^{\infty}J_0(\rho\alpha)
e^{-|z_e-z_h|\alpha}d\alpha
\end{equation}
and after solving the $\rho$ integral, it yields
\begin{equation}
\label{i-final}
\int_0^{\infty} d\alpha(\frac{\alpha^2}{4g_B})^me^{\frac{-\alpha^2}{4g_B}}
\int_{-\infty}^{\infty} dz_edz_h\psi_{\nu'_e}\psi_{\nu'_h}
e^{-|z_e-z_h|\alpha}\psi_{\nu_e}\psi_{\nu_h}.
\end{equation}

The $z_e$ and $z_h$ integrals can be solved analytically.
The evaluation of these integrals
is cumbersome due to the large number of terms, resulting
from the five different regions of the potential.
The $z_e$ and $z_h$ integrals contain both,
decoupled terms in which the $z_e$ and $z_h$ integrals
are independent of each other, and coupled terms where
the integration limits of the $z_h$ integral contains $z_e$.
The remaining $\alpha$ integral must be calculated
numerically. 

Diagonalizing the system of equations resulting for
the coefficients in the expansion (\ref{expansion}),
in a truncated basis,
we obtained the energies and wave-functions for the
first excitonic levels.
Evaluating the oscillator strength
\begin{equation}
\label{os}
f_n=\frac{|\int dz\Psi^{exc}_n(\rho=0,z_e=z,z_h=z)|^2|\vec{\epsilon}
\cdot\vec{p}_{eh}|^2}{m_0E_n}
\end{equation}
we obtained the optical absorption spectra.

\section{Results\protect\\}

We calculated the excitonic energy levels and
the magneto-optical absorption spectra for
symmetric and asymmetric double quantum wells,
composed by $GaAs$ wells and $Al_xGa_{1-x}As$ barriers ($x=0.3$).
The band gap used in our calculations is given by
$E_g(x)=1.52+1.36x+0.22x^2$.
The band gap offset considered
was a sixty percent for the conduction band and a
forty percent for the valence band.
For comparison purposes in the symmetric case,
we used the same simplified masses of Ref.\ \onlinecite{cen},
for all five regions in the double quantum well.
The electron mass considered was $m_e=0.067m_0$,
the $x-y$ plane heavy hole mass, $m_{hh,x-y}=0.1m_0$,
the $z-axis$ heavy hole mass $m_{hh,z}=0.45m_0$ and the light hole masses
$m_{lh,x-y}=0.2m_0$, $m_{lh,z}=0.08m_0$, which
corresponds to $\gamma_1=7.36$ and $\gamma_2=2.57$.
We considered a dielectric constant $k_0=12.5$.

It was enough to consider in our calculations
a truncated basis composed of twelve Landau wave-functions,
four electronic wave-functions and four heavy hole
wave-functions.

\subsection{Symmetric double quantum well\protect\\}
\label{sec:Symmetric}

In this section we present our results for the excitonic
energy levels in symmetric double quantum wells
as a function of wells and barrier widths,
ranging from zero to large values,
in a $10\ Tesla$ magnetic field.
Also, we present the optical absorption spectra, including
broadening for each possible optical transition.

Fig.\ \ref{binding}a shows  the binding energy
of the ground state, for a $25$\ \AA\  barrier
and for  both wells widths ranging from $0$ to $200$\ \AA\ .
For zero well width, the exciton is tridimensional
and it corresponds to bulk $Al_{.3}Ga_{.7}As$.
For non-zero well width, the exciton is localized within
the wells, reaching a maximum binding energy at a certain well
width which corresponds to a maximum two-dimensional character.
For further increase in the wells width,
the binding energy decreases and the exciton
evolves towards a tridimensional $GaAs$ exciton for thick wells.
This behavior of the ground state compares very
well with the results of Ref.\ \onlinecite{cen},
although our energies are $0.5\ meV$ lower.

Fig.\ \ref{binding}b shows the ground state binding energy
for wells width fixed in $100$\ \AA\ ,
and barrier width ranging from  $0$ to $100$\ \AA\ . 
As the barrier width is varied from zero to infinity,
we move from the case of a single quantum well (width: $Lw1+Lw2$),
towards two independent quantum wells of width $Lw1$ and $Lw2$.
It is known that when the system size decreases,
the exciton binding energy increases, accordingly
the binding energy starts in a value which corresponds to a
single quantum well of width $Lw1+Lw2$
and ends in a larger value that corresponds to
each decoupled quantum well.
The binding energy decreases for thinner barriers, because
the wave-functions have a large amplitude within the barrier,
which is equivalent to a single quantum well wider than $Lw1+Lw2$.
In this case we obtain an excellent agreement with
Ref.\ \onlinecite{cen}.

Fig.\ \ref{sym-lb}a shows the ground state and
higher excitonic levels in a symmetric double quantum well,
for $100$\ \AA\  wells and barrier width ranging
from $0$ to $50$\ \AA\ .
It can be seen that the second energy level
(optically prohibited in the symmetric case),
gets closer to the ground state for wider barriers.
This is a typical behavior, when we go from a situation of
coupled quantum wells towards one of two decoupled wells
having degenerate energies. In this case
the Coulomb potential do not mix states with
different symmetry under simultaneous interchange of
$z_e$ by $-z_e$ and $z_h$ by $-z_h$.
In terms of our base wave-functions, the first two excitonic
states evolve from $e_1h_1$ (first electron - first hole, exciton)
and $e_1h_2$ (first electron - second hole, exciton),
towards $e_1h_1+e_2h_2$
and $e_1h_2+e_2h_1$ as the barrier get wider.
This behavior was noted (without magnetic field) in
Ref.\ \onlinecite{dignam} 
and corresponds to the ground state evolving towards $e_Lh_L+e_Rh_R$,
in terms of a basis of single quantum well wave-functions,
corresponding to the electron (or hole) localized in the left well $L$
or the right well $R$.
This classification in terms of $L$ and $R$ states, will be
useful in explaining the optical absorption spectra  for thick barriers.

In Fig.\ \ref{sym-lb}b, we show the optical absorption
spectra for $100$\ \AA\  wells,
with the central barrier ranging from $0$ to $30$\ \AA\ .
The state $e_2h_2$ evolves toward $e_1h_1-e_2h_2$
getting close to the state $e_1h_1+e_2h_2$.
This state disappear from the spectra,
because is evolving towards $e_Rh_L+e_Lh_R$,
where the coupling between electron and hole wave-functions
decreases, as the electron and the hole are localized in separated wells.
The finite energy separation between these states for wide barriers,
is a consequence of the small binding energy for the $e_1h_1-e_2h_2$
state, and corresponds to the binding energy for the $e_1h_1+e_2h_2$
state for very large barriers.
For thin barriers, one small peak between $e_1h_1$ and $e_2h_2$ ,
corresponding to $e_1h_3$, can not be distinguished.
The other peaks correspond to higher levels and the shoulder
in $e_2h_2$, which evolves growing in size, corresponds to
a Landau level of $e_1h_1$.

The states involved in the anti-crossing (produced for a
$10$\ \AA\  barrier) have a very small oscillator strength,
compared to the $e_2h_2$ state. Because of this, the anti-crossing
manifest itself only as small distortion, in the evolution from
$e_2h_2$ towards $e_1h_1-e_2h_2$, which finally disappear from
the spectra for thick barriers, as mentioned previously.

Fig.\ \ref{sym-lw}a shows the ground and higher
excitonic states (with respect to the first band to band transition)
for a symmetric double quantum well,
with a central barrier of $25$\ \AA\ 
and both wells ranging from $0$ to $200$\ \AA\ .
The higher levels show an anticrossing region
involving the states $e_2h_1$ ($e_2h_2$) and a Landau level.
The states $e_1h_1$ and $e_1h_2$ evolve
towards $e_1h_1+e_2h_2$ y $e_1h_2+e_2h_1$.

Fig.\ \ref{sym-lw}b shows the calculated optical absorption
for the previous case.
The anticrossing region (for $45$\ \AA\ wells) manifests as
a reinforcement in the spectral lines,
caused by the
coincidence of the Landau energy and the energy separation
between the $e_1h_1$ and $e_2h_2$ states.
It can be seen as the $e_2h_2$ state get close to the $e_1h_1$ state
and loose probability for wide wells.

Previous works only included the
coupling between levels, necessary for having the right energy behavior
for the excitonic wave-function for thick barriers,
and the general belief was that coupling between others levels
is of no importance.
In our calculation is clear that although this is true for
almost all wells and barrier widths, there are regions where
another off-diagonal terms must be included, for explaining
the anti-crossings that we obtained between higher levels
(see Fig.\ \ref{sym-lb}a).

\subsection{Asymmetric double quantum well\protect\\}
\label{sec:Asymmetric}

When going from the symmetric case towards the
asymmetric one, states prohibited by the quantum
well symmetry can appear in the optical absorption
spectra. These states have a small oscillator strength,
near the symmetric case, which increases when the
$z$\ potential is going far from symmetry.

In asymmetric double quantum wells,
the excitonic ground state wave-function is mostly
localized in the wider well.
For $Lw1$ smaller than $Lw2$ the ground state binding
energy corresponds to an exciton localized in the right well.
As $Lw1$ get closer to $Lw2$, the exciton begin to be
present in the left well and the binding energy decreases.
When the two wells are of the same width, the exciton is
localized in both wells (symmetric case)
and the binding energy reach a local minima.
For $Lw1$ bigger than $Lw2$, the exciton
exciton is localized in the left well and the
binding energy increases.
After this, the binding energy reach a maxima and decreases
towards the tridimensional limit of a very thick left well.

Fig.\ \ref{asym-lw1}a shows the ground and higher excitonic
states, for a $100$\ \AA\  right well, a $35$\ \AA\  barrier
and a $10\ Tesla$ magnetic field,
for the left well ranging from $60$\ \AA\  to $100$\ \AA\ .
The ground state remains fixed in energy and corresponds to
an exciton localized in the right well (the ground state
binding energy behavior, explained previously, can not be
observed in the scale of Fig.\ \ref{asym-lw1}).
The second excitonic state, prohibited and degenerate
in energy with the ground state (for thick barriers) in the symmetric case,
separates from the ground state when the left well width
decreases. For a $85$\ \AA\  left well, this state interacts
with the third state, producing an anti-crossing.
In the anticrossing region, these states lose their identity,
transforming into $e_1h_2+e_2h_2$ and $e_1h_2-e_2h_2$.
From analyzing the energies and wave-functions in our basis states,
we see that this anti-crossing is not originated by a crossing
between electron (or hole) levels of the right and left wells
(see Ref.\ \onlinecite{ferreira1990}),
as for almost all left well widths of Fig.\ \ref{asym-lw1}a,
the electron (hole) levels $e_1,e_3$ ($h_1,h_3$) 
are already localized in the right well and
$e_2,e_4$ ($h_2,h_4$) in the left well.
These anti-crossings, are a consequence of the
strong interaction between excitonic states,
when the levels corresponding to the direct exciton
$e_2h_2$ ($e_Lh_L$) and indirect exciton $e_1h_2$ ($e_Rh_L$),
get close in energy.
The different behavior in these energy levels,
when one of the wells width is varied,
is originated by the different spatial
character of excitons localized in only one well (direct)
and excitons where electron and hole are localized in
different wells (indirect).
As can be seen from Fig.\ \ref{asym-lw1}b,
both excitonic states have a large oscillator strength 
in the anti-crossing region, and because the
energy separation is approximately $4\ meV$ between them,
this anti-crossing could be observed experimentally.

In Fig.\ \ref{asym-lb-B}a we show the excitonic states
for a $85$\ \AA\ left well and a $100$\ \AA\ right well
in a $10\ Tesla$ magnetic field,
as a function of the barrier width.
It is clear the way the anti-crossing states evolve and
get separated in energy as the barrier is decreased.
When these energy levels go far from each other,
the probability of being created optically decreases.
This would make difficult to obtain experimentally this
anti-crossing, for barriers smaller than $35$\ \AA\ .
For wider barriers, the direct exciton $e_2h_2$
has a lower energy than the indirect exciton $e_1h_2$,
as a consequence of the large binding energy of $e_2h_2$, relative
to $e_1h_2$.
These different binding energies are a consequence of
the large spatial separation of
the wave-functions corresponding to $e_1$ and $h_2$.

These anti-crossings has not be obtained previously.
We think that it is important to carry out an experimental
study, of the correlation in the optical absorption spectrum,
for the energies of the second and third excitonic
states, in the anti-crossing region.
Also could be interesting, to study the possible relationship
of the states $e_1h_2+e_2h_2$ and $e_1h_2-e_2h_2$ 
with charged excitons \cite{shields,finkelstein,lelong},
because these states involve one particle wave-functions,
corresponding to two electrons and one hole states.

\subsection{Magnetic field effects\protect\\}
\label{sec:Magnetic}

Increasing the magnetic field in a symmetric or
asymmetric double quantum well, produces a
shift in the excitonic energy levels towards higher energies
and an increase in the binding energy.
The increase in the binding energy is a consequence of two effects:
First, the wave-function confinement in the $x-y$ plane
produces a stronger interaction in the $z-axis$, as the electron
and hole wave-functions penetrates more in the barrier.
Second, the binding between electron and hole in the
plane increases, because their wave-functions are confined
to a smaller region.

The large binding between electron and hole
obtained when increasing the applied magnetic field, is similar
to a change in the barrier and well width. This can be used
to study these systems in regions of interest, without
the need for the growth of many different samples.

In Fig.\ \ref{asym-lb-B}b, we show the magnetic field effects
for the asymmetric double quantum well, in the anti-crossing
region of Fig.\ \ref{asym-lb-B}a.
When the magnetic field ranges from $0$ to $40\ Tesla$,
the energy levels are shifted towards a bigger energy and
the binding is stronger.
The behavior for the indirect excitonic levels is
similar to decreasing the barrier width in the
anti-crossing region.
The direct exciton binding energy also increases,
as a consequence of the wave-function localization in the $x-y$ plane.
Because the direct exciton has a larger binding energy than
the indirect exciton for large magnetic fields, this level goes below the
indirect exciton level for strong magnetic fields.

\section{Conclusion\protect\\}
\label{sec:Conclusion}

In this work we studied the energies 
and oscillator strength, for several excitonic $\bf S$ levels
in a double quantum well, in a magnetic field pointing in
the growth direction $z$.
We calculated the excitonic binding energy
and optical absorption for type $I$
symmetric and asymmetric double quantum wells,
as a function of barrier and well (or wells) widths.

In our method we used an orthogonal basis
and we do not employed a variational method.
Specific calculations for the ground state excitonic binding energy,
in symmetric double quantum wells, are in excellent
agreement with the work of Cen-Bajaj\cite{cen}.
In contrast to the Dignam-Sipe\cite{dignam} work,
we used basis functions of single particle solutions
of the double quantum well, and our method is valid for
all barrier width.

To our understanding, this is the first theoretical calculation,
in a symmetric and asymmetric double quantum well, that is
valid for all barrier widths.
This allow us to study these systems, from the case of a single
quantum well (null barrier) to the case where both wells get
decoupled.

We do not know another theoretical calculation that predict
the anti-crossings between excitonic levels, that we obtain
as a consequence of including in our method all the coupling
between our basis wave-functions.

\acknowledgments
We want to acknowledge the hospitality of the people in the
electronic department in Glasgow University, specially to
C. Sotomayor, D. Hutchings and J. Arnold.
We also want to acknowledge a Conicyt-British Council grant
for a research student-ship given to Francisco Vera,
and to G. Fuster for many discussions on this topic.
This work was supported by the Fondecyt projects
1950190,1970119 and a Conicyt doctoral fellowship.

\begin{figure}
\caption{Potential profile in the $z$ direction for electrons and holes.
}
\label{pot-z}
\end{figure}

\begin{figure}
\caption{Exciton binding energy for a symmetric double quantum
         well ($B=10\ Tesla$).
	(a) Plotting as a function of the width of the
	 wells, for a barrier of $2.5\ nm$. (b) As a function
	 of the width of the barrier, for wells of $10\ nm$.}
\label{binding}
\end{figure}

\begin{figure}
\caption{Exciton energy levels for a symmetric double quantum
	 well as a function of the barrier width, for wells of $10\ nm$
	 ($B=10\ Tesla$).
	 (a) Evolution of the excitonic states, showing the most
	 important base states present in the  exciton wave-function
       (ij means $e_ih_j$).
	 (b) Calculated optical absorption for these states, each curve
	 represent a different barrier width showed by the $y$\ axis,
	 the peaks show the optical absorption refered to the
	 bottom of each curve.}
\label{sym-lb}
\end{figure}

\begin{figure}
\caption{Exciton energy levels for a symmetric double quantum
	 well as a function of the wells width, for a barrier of $2.5\ nm$
	 ($B=10\ Tesla$).
	 (a) Evolution of the excitonic states.
	 For clarity in representing these states,
	 we subtracted the first energy level in this system when
	 the Coulomb potential is not taken into account.
	 (b) Calculated optical absorption for these states.}
\label{sym-lw}
\end{figure}

\begin{figure}
\caption{Exciton energy levels for an asymmetric double quantum
	 well as a function of the left well width, for a barrier
	 of $3.5\ nm$ and a right well width of $10\ nm$ ($B=10\ Tesla$).
	 (a) Evolution of the excitonic states.
	 (b) Calculated optical absorption for these states.}
\label{asym-lw1}
\end{figure}

\begin{figure}
\caption{Exciton energy levels for an asymmetric double quantum
	 well, with a left well width of $8.5\ nm$ and right well
	 width of $10\ nm$.
	 (a) As a function of the barrier width ($B=10\ Tesla$).
	 (b) As a function of the magnetic field for a barrier
	 of $3.5\ nm$.
       (numbers enclosed in parenthesis indicates a small
       contribution of this states, to the exciton wave-function) }
\label{asym-lb-B}
\end{figure}

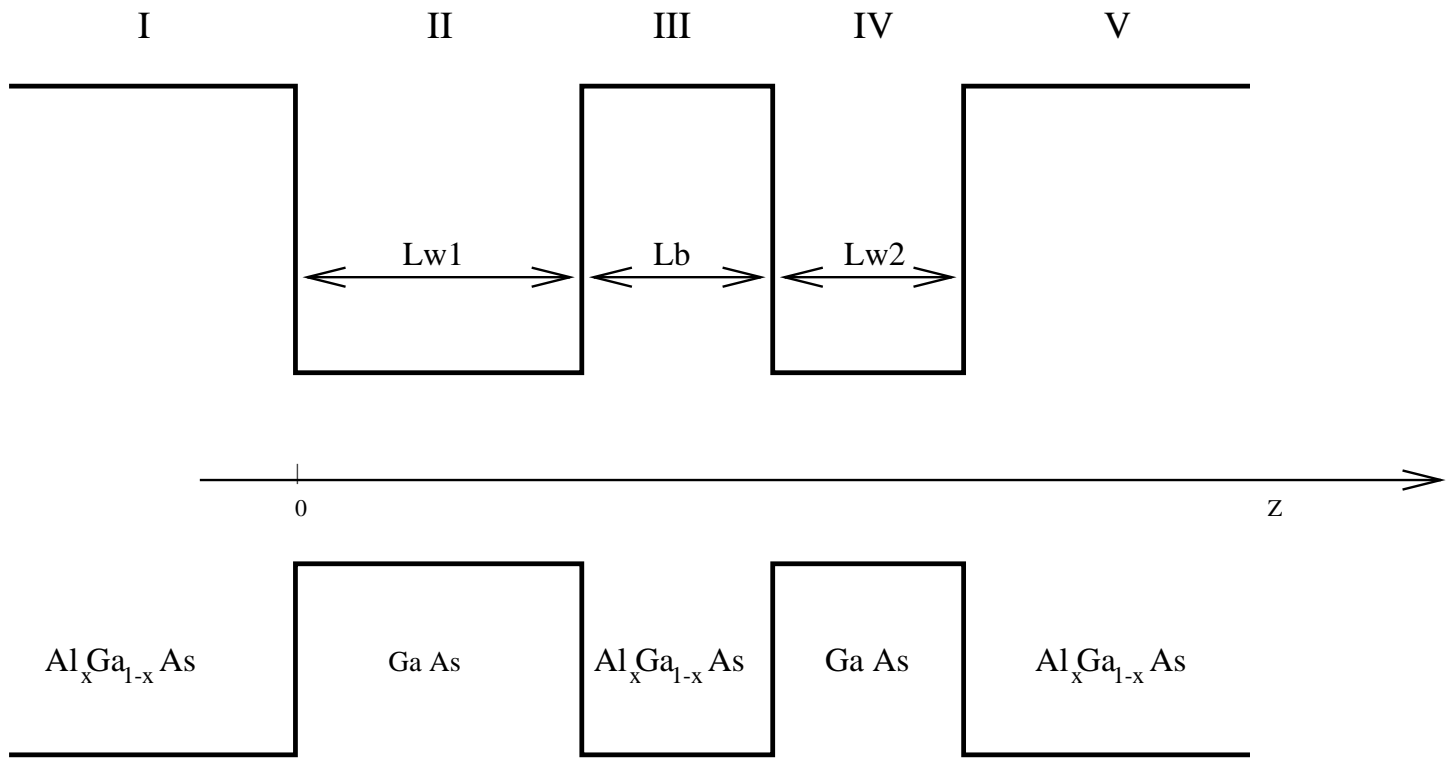
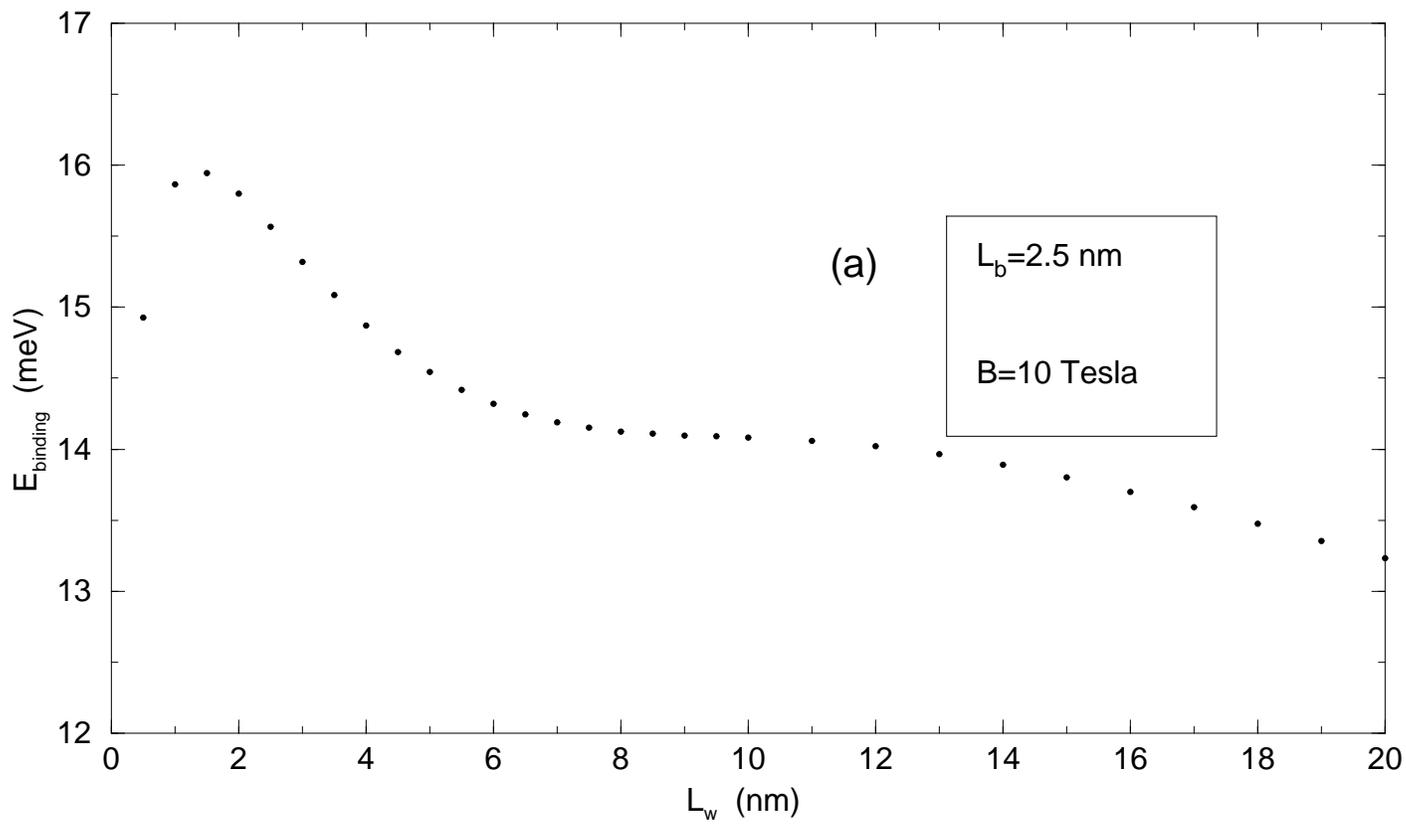
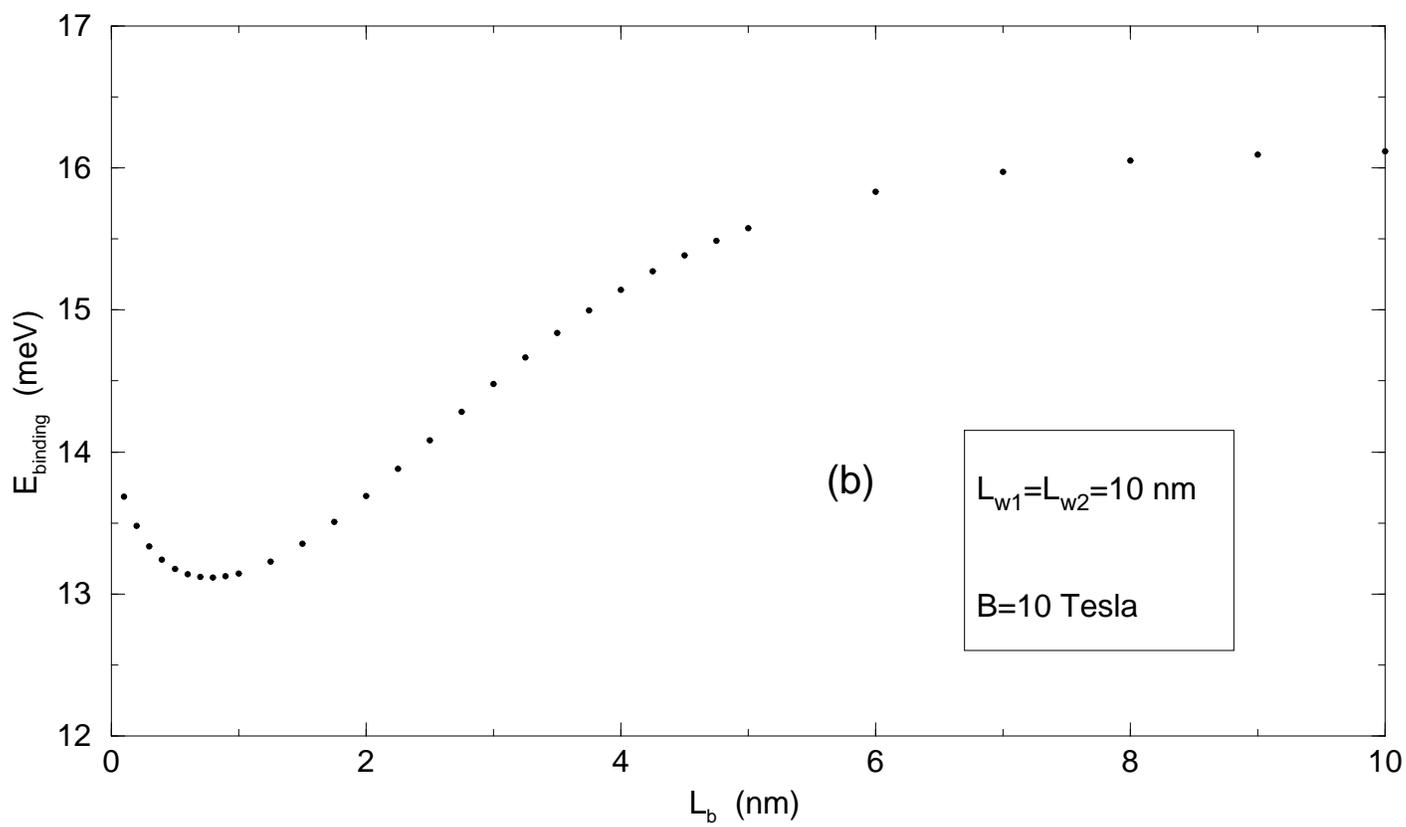
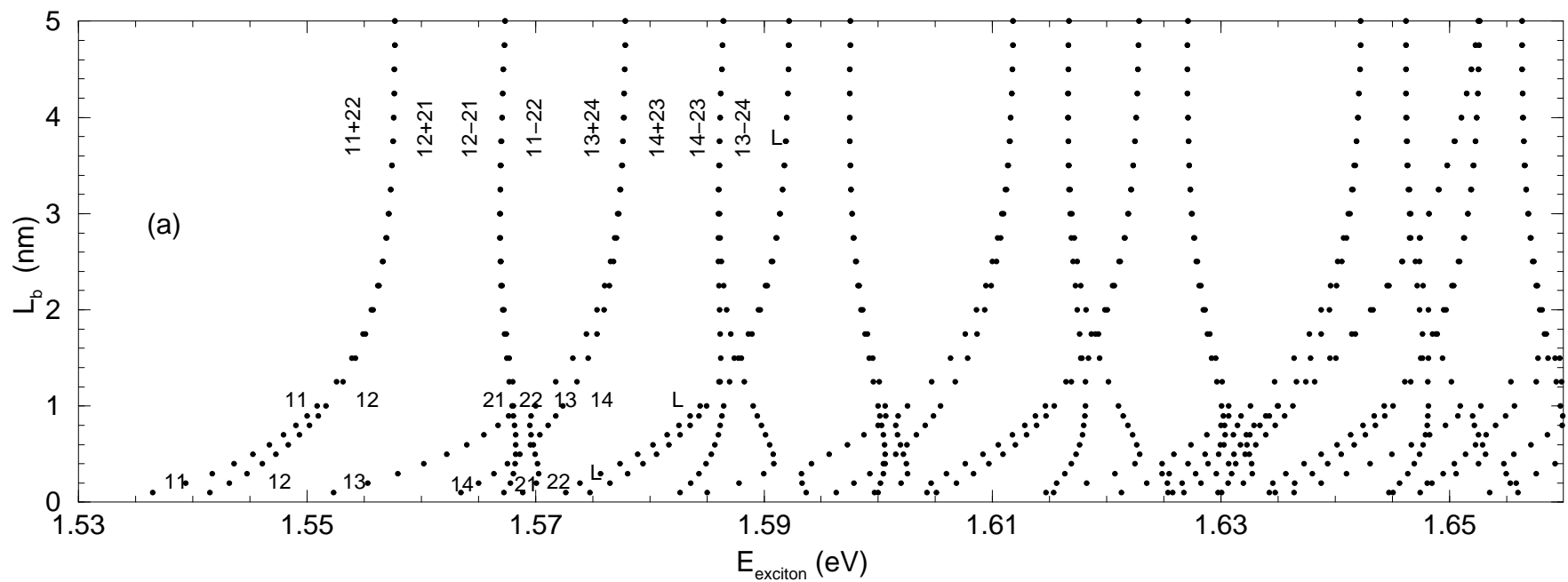
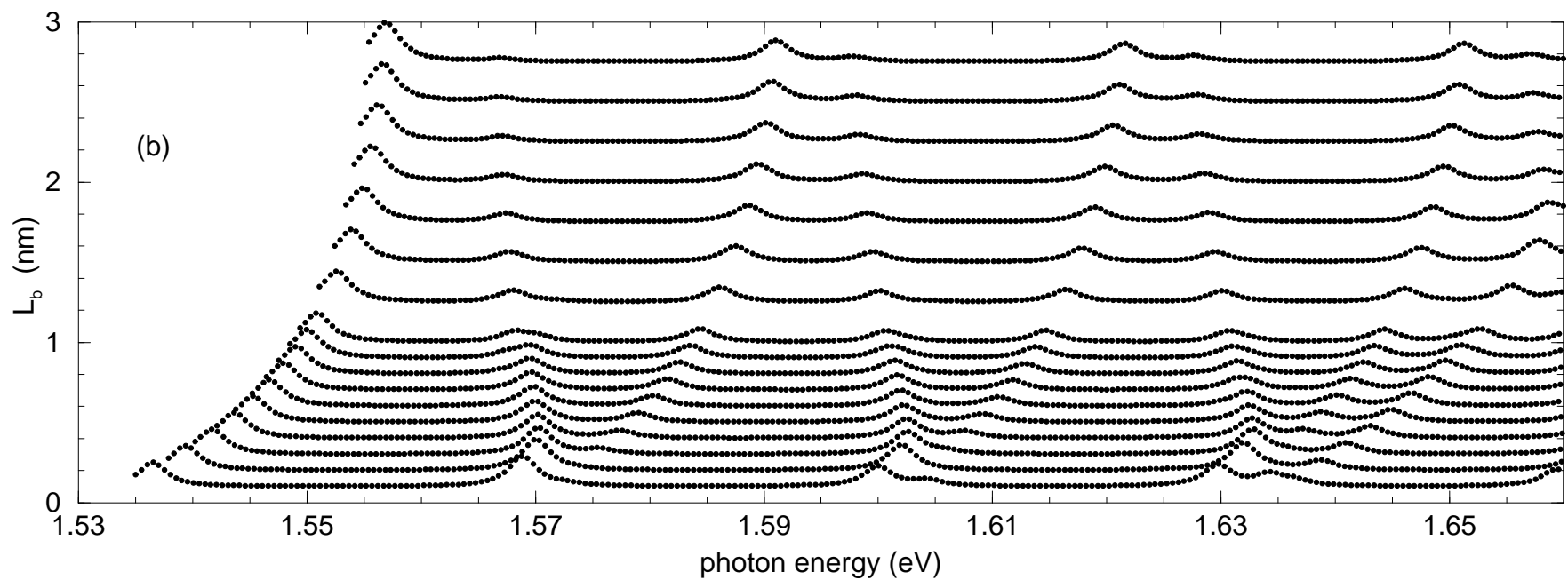
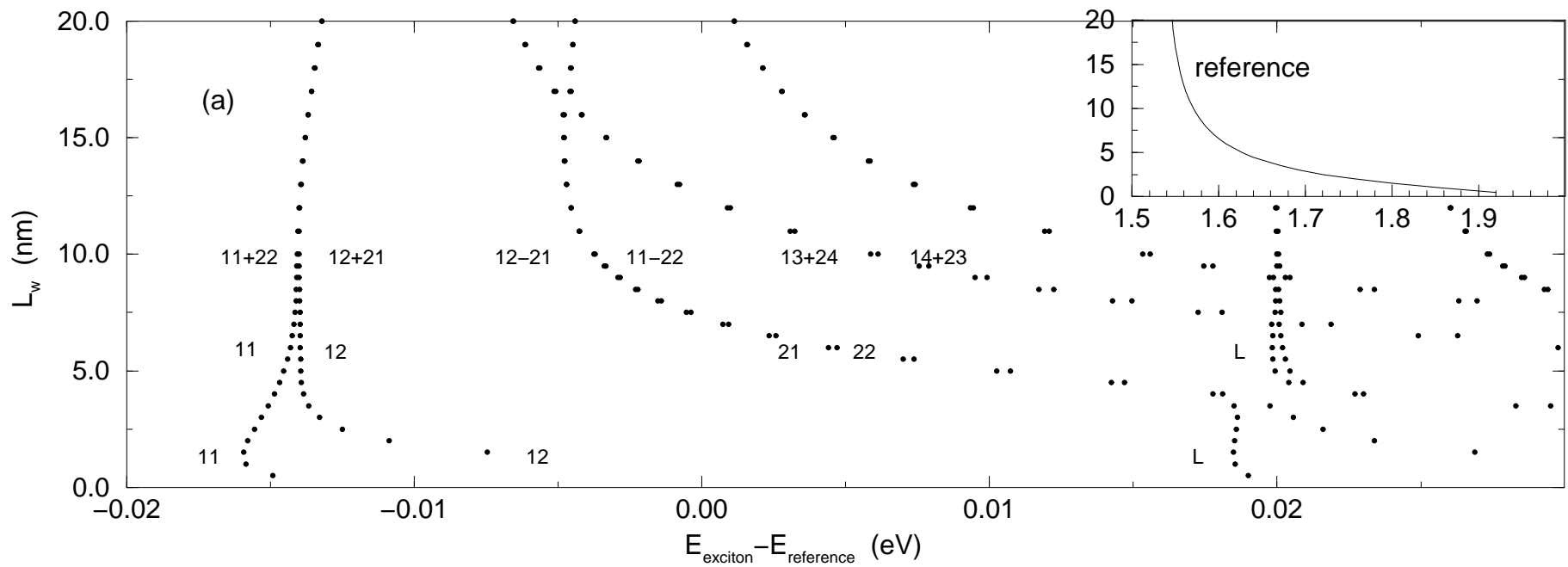
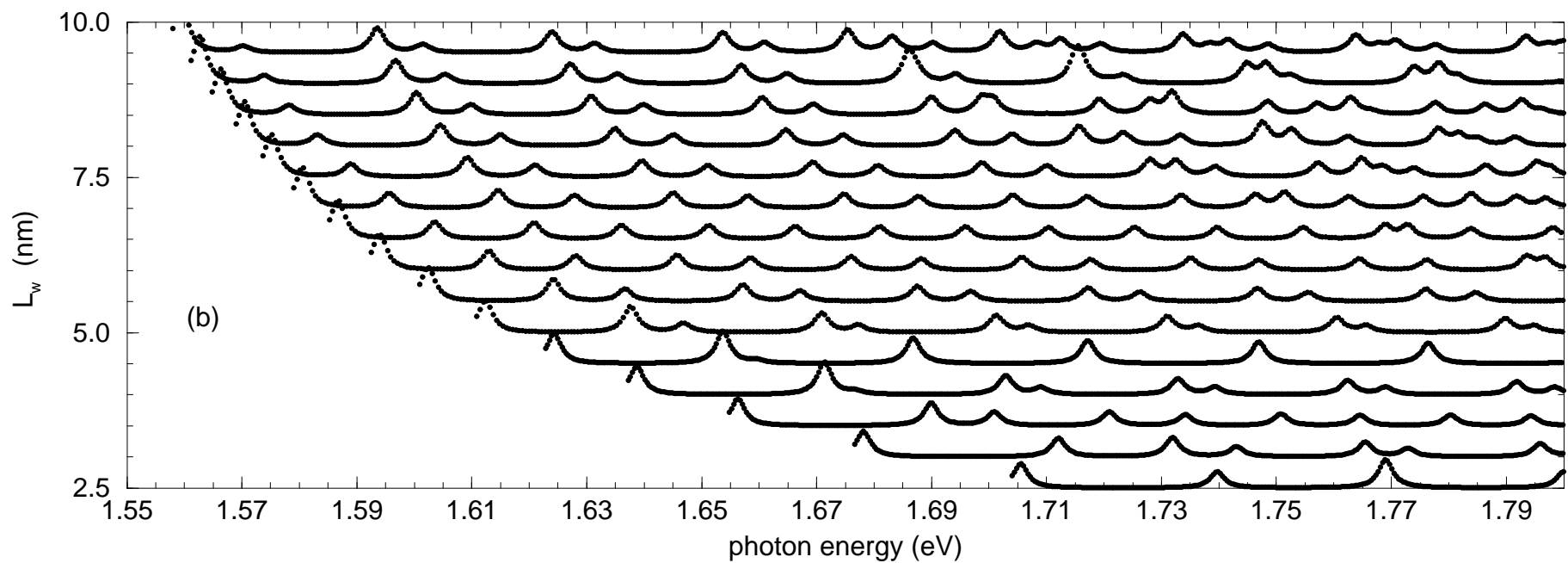
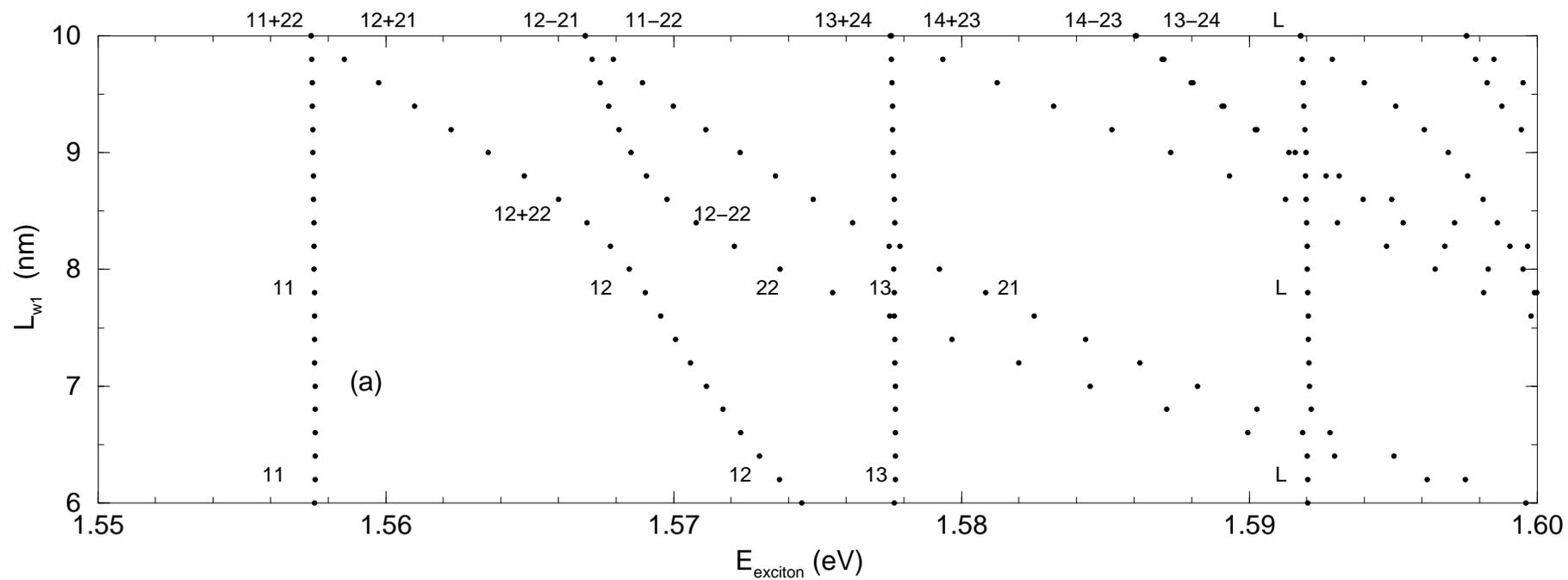
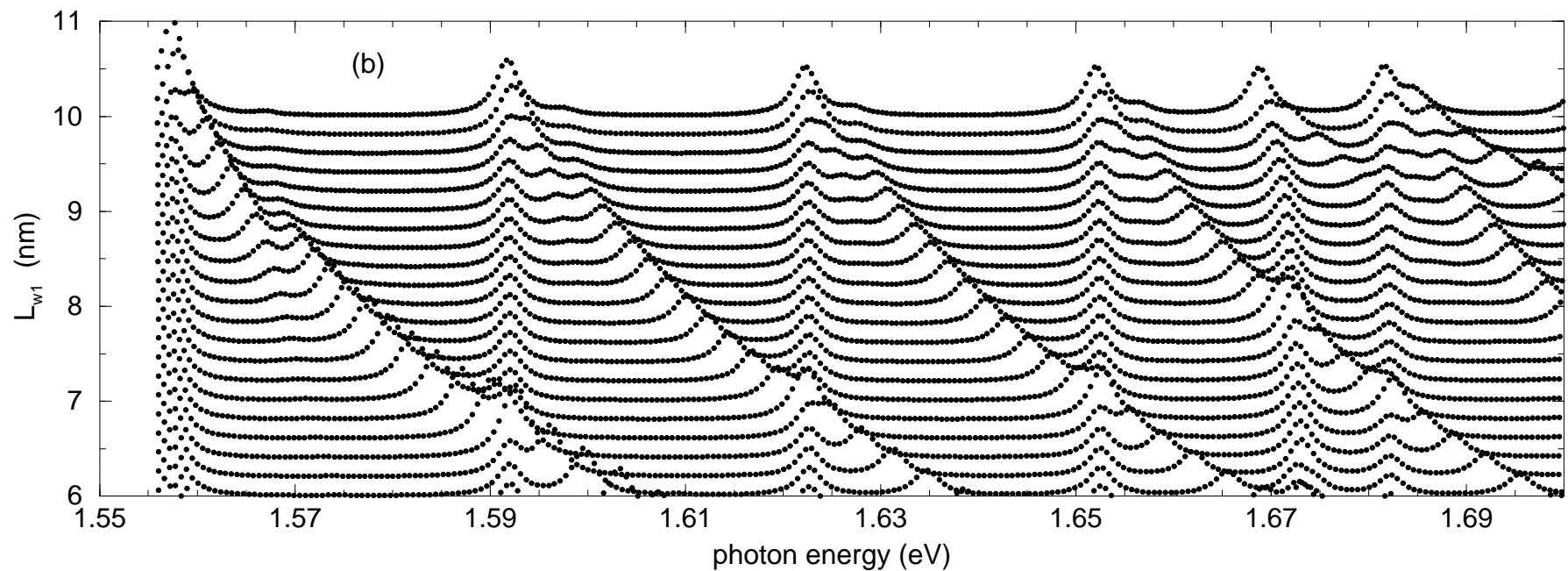
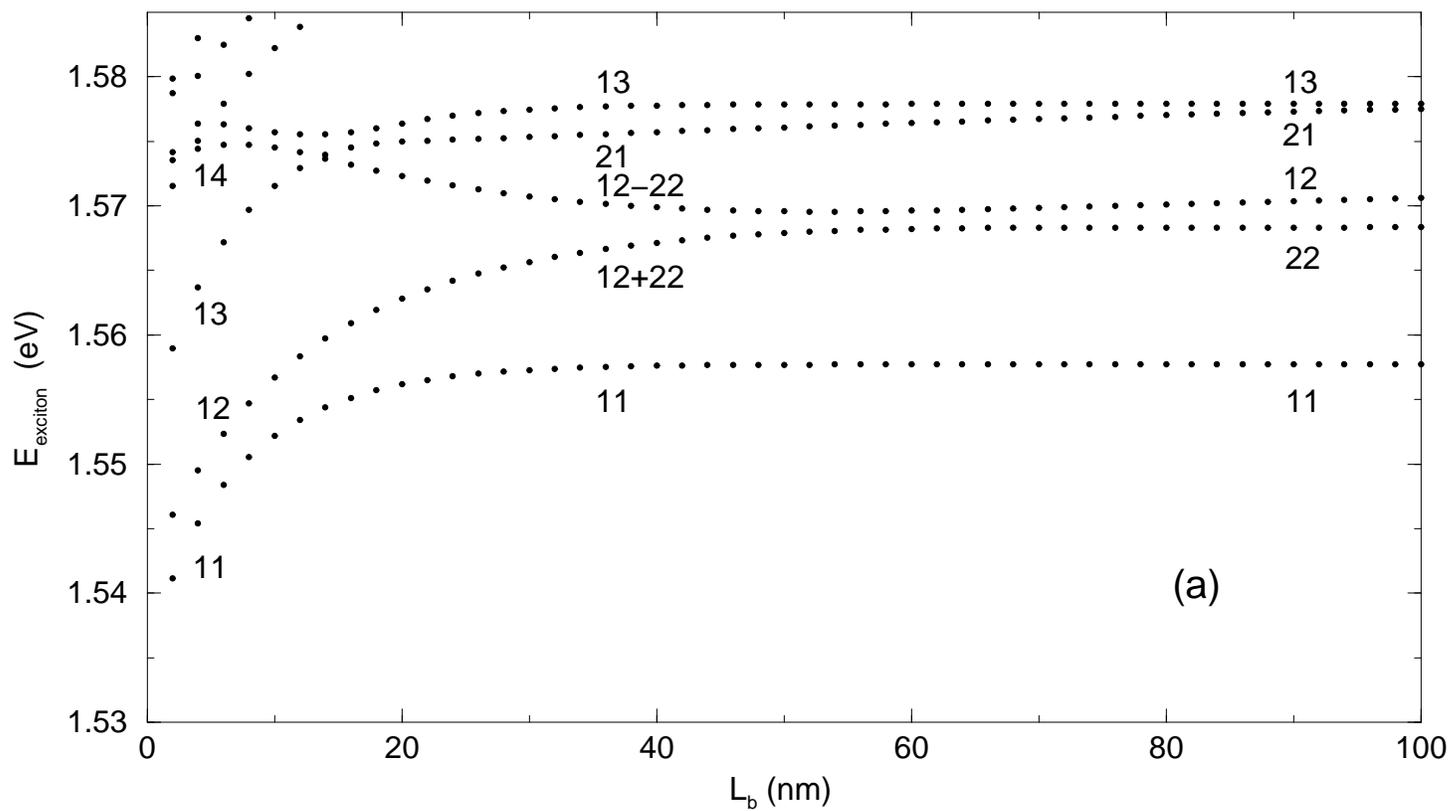
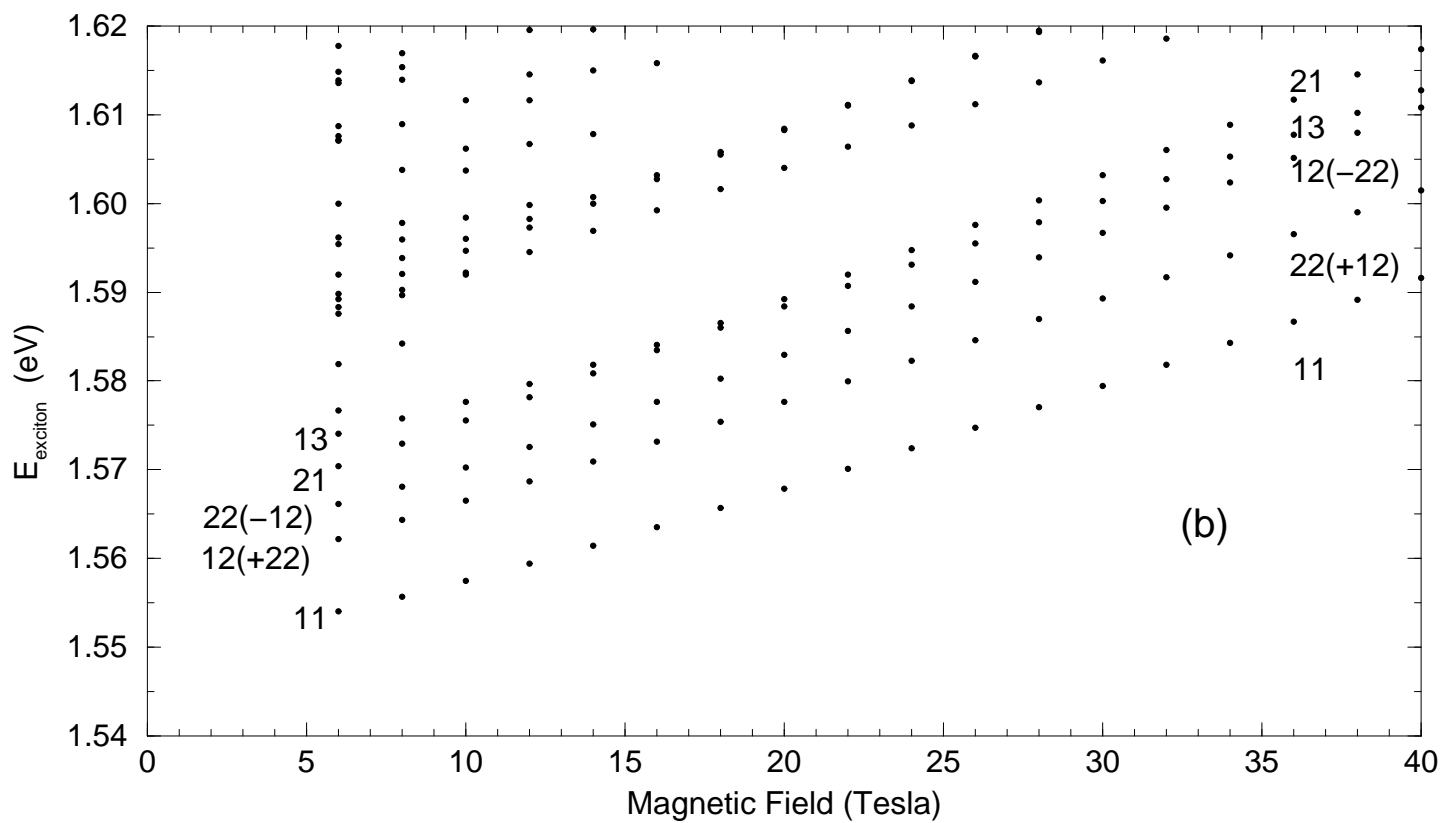
\end{document}